\documentclass[twocolumn]{aastex631}
\usepackage{float}
\usepackage{amsmath}
\usepackage{rotating}
\usepackage{amssymb}
\usepackage{hyperref}
\usepackage{ulem,color}
\usepackage{graphicx}
\usepackage{xcolor}

\def\gtrsim{\mathrel{\hbox{\rlap{\hbox{\lower4pt\hbox{$\sim$}}}\hbox{$>$}}}}
\def\lesssim{\mathrel{\hbox{\rlap{\hbox{\lower4pt\hbox{$\sim$}}}\hbox{$<$}}}}
\def\gtrsim{\mathrel{\hbox{\rlap{\hbox{\lower4pt\hbox{$\sim$}}}\hbox{$>$}}}}

\begin{document}

\newcommand {\rosat} {{\it ROSAT}}
\newcommand {\einstein} {{\it Einstein}}
\newcommand {\exosat} {{\it EXOSAT}}
\newcommand {\asca} {{\it ASCA}}
\newcommand {\ginga} {{\it Ginga}}
\newcommand {\tenma} {{\it Tenma}}
\newcommand {\sax} {{\it BeppoSAX}}
\newcommand {\hst} {\textsl{HST}}
\newcommand {\xmm} {\textsl{XMM-Newton}}
\newcommand {\chandra} {\textsl{Chandra}}
\newcommand {\swift} {\textsl{Swift}}
\newcommand {\rxte} {\textsl{RXTE}}
\newcommand {\fermi} {\textsl{Fermi}}
\newcommand {\nustar} {\textsl{NuSTAR}}
\newcommand {\suzaku} {\textsl{Suzaku}}
\newcommand {\integral} {\textsl{INTEGRAL}}

\newcommand{\nk}[1]{\textcolor{purple}{#1}}

\def\chan{{\sl Chandra\ }}
\newcommand{\hl}[1]{{\color{red}\bf\large #1}}
\newcommand{\jh}[1]{{\color{red} #1}}
\def \src {LS5039}

\title{X-ray imaging observations  of  the high-mass $\gamma$-ray  binary HESS~J0632+057}

\author{Kargaltsev O.}
\affiliation{The George Washington University, Department of Physics, 725 21st St NW, Washington, DC  20052}
\author{Klingler, N.~J.}
\affiliation{Department of Astronomy \& Astrophysics, The Pennsylvania State University, 525 Davey Laboratory, University Park, PA, 16802, USA}
\affiliation{Astrophysics Science Division, NASA Goddard Space Flight Center, Greenbelt, MD, 20771, USA}
\affiliation{Center for Space Sciences and Technology, University of Maryland, Baltimore County, Baltimore, MD, 21250, USA}
\author{Hare, J.}
\affiliation{Astrophysics Science Division, NASA Goddard Space Flight Center, Greenbelt, MD, 20771, USA}
\altaffiliation{NASA Postdoctoral Program Fellow}
\author{Volkov, I.}
\affiliation{The George Washington University, Department of Physics, 725 21st St NW, Washington, DC  20052}

\begin{abstract}

The {\sl Chandra X-ray Observatory} {\sl(CXO)} imaged the high-mass $\gamma$-ray binary (HMGB) HESS J0632+057  with the Advanced CCD Imaging Spectrometer (ACIS).  We analyzed the {\sl CXO} data together with 967 ks of archival {\sl Swift}-XRT observations.   On arcsecond scales    we find a hint of asymmetric extended emission.      On arcminute scales,     a region  of extended  emission (``the blob''), located  $\approx5.5'$ east of the binary, is seen in both the {\sl CXO}-ACIS and the {\sl Swift}-XRT images. The blob does not    have a counterpart in the radio, NIR, IR, or optical images.     The ACIS spectrum of the blob fits either an absorbed power-law model with $\Gamma\simeq2.6$, or a thermal plasma model with $kT\simeq 3$ keV.     Since the blob's $N_{\rm H}$ is significantly larger than that of the binary we conclude that the blob and binary are not directly related.      The somewhat larger very deep XRT image suggests that the binary may be located within a shell (or cavity).     The four ACIS spectra taken within the  $\sim$20 day interval  near the light curve minimum  suggests that the $N_{\rm H}$ is varying on timescales of days, possibly due to the inhomogeneous circumbinary environment.  The XRT spectra extracted from the wider orbital phase intervals  support significant changes in  $N_{\rm H}$ near the light curve maximum/minimum, which may be responsible for the substantial systematic residuals seen near 1 keV, and provide tentative evidence for a Fe line at 6.4 keV. We find no significant periodic signal in the ACIS data up to 0.156 Hz.

\end{abstract}


\section{INTRODUCTION}

High-mass $\gamma$-ray binaries  (HMGBs) are rare high-mass binaries containing O or Be type stars and a compact object. To date there are 8 HMGBs firmly known with one  of them being in LMC and the rest in our own Galaxy. 
Two Galactic HMGBs with the widest, most elliptical orbits host known radio pulsars (B1269--63 and TeV J2032+4130).  
In other HMGBs, the nature of the compact objects is unknown with both NSs and BHs being possibilities (see \citealt{2020arXiv200603615C} and references therein).   
    HMGBs with young pulsars 
     represent an extreme form  of colliding wind binaries 
     allowing one to probe properties of relativistic pulsar winds. HMGBs with BHs  accreting  from the companion wind can be used to probe  accretion flow physics in the regime where the accretion rate is low and varies periodically with time.

HMGB HESS J0632+057 (hereafter J0632) discovered by \cite{Hinton2009},
shows variable emission in TeV $\gamma$-rays as seen with  with VERITAS, H.E.S.S., HAWC, and MAGIC observatories (see \citealt{2021arXiv210911894A} and references therein).
ESS~J0632 is projected near the edge of the star-forming region of the Rosette Nebula.
The photometric distance, obtained from the SED fits to the massive star spectrum is  $1.1-1.7$ kpc \citep{Aragona2010}, is consistent with the distance to the Rosette Nebula star-forming region ($\sim1.4$ kpc; e.g., \citealt{Gahm2007}). 
The binary is located at 
$d=1773^{+93}_{-88}$ pc according to  
Gaia EDR3 distance catalog \citep{Bailer-Jones2021}. 
Throughout this paper we adopt the 1.5 kpc distance commonly used in the literature. 
Although the most recent estimate of the distance based on Gaia EDR3 is somewhat larger, it can be affected by unaccounted for systematics  
due to the uncorrected effects of binary motion impacting the astrometry.

\begin{table*}
\caption[]{   {\sl CXO} ACIS observations of  J0632.
} \vspace{-0.5cm}
\begin{center}
\begin{tabular}{ccccccc}
\tableline\tableline ObsID & Start Time (MJD)  & Exposure$^{\rm a}$ (s) & $T_{\rm span}$ (s) & Orbital Phase$^{\rm b}$ & Cts$^{\rm c}$  \\
\tableline
& &{\sl CXO} CC-mode  &&\\
\hline 
13237 & 55605.88101852 & 39,920 & 40,074 & 0.362[0.391]  &  10,703\\
& &{\sl CXO} imaging& & \\
\hline 
& & ~~~~ {\sl CXO} imaging Epoch 1  &&\\
\hline 
20269 & 58152.7673 & 33,630 &  34,075 & 0.402 [0.528] &  463 \\
20950 & 58153.5007 & 45,476 & 46,077 & 0.404 [0.530]  & 539 \\
\hline 
& & ~~~~{\sl CXO} imaging Epoch 2  &&\\
\hline 
20975 & 58168.6285 &41,523  & 42,071 & 0.452 [0.579] & 1,157 \\
20974 & 58170.0665 & 31,658 & 32,070 & 0.456 [0.583] & 1,043 \\
   \tableline
\end{tabular}
\end{center}
$^{\rm a}$ -- Live time. Note, for the first two observations the scientific exposure for J0632 is reduced by a factor of 1.5 due to it dithering across the chip gap. $^{\rm b}$ -- The binary phases correspond to the start time of the corresponding observation. They are calculated according to M+19 
   ephemeris ($P_b=316.8^{+2.6}_{-1.4}$ days)  and  Moritani et al.\ (2018)
   ephemeris ($P_b=313^{+11}_{-8}$ days), respectively.  Zero phase  $\phi=0$ in both cases is chosen to be 
   54857 MJD. $^{\rm c}$ -- Number of photons in 0.5-8 keV within the $r=3''$. Note that the count rates differ so much because of the source being dithered across the chip gap during the Epoch 1 observations.
\end{table*}

The optical counterpart of J0632 is a massive B0pe 
star MWC 148 (HD 259440 = LS VI +05 11) which shows  
brightness variations 
at the level of 1\%-1.3\% in the optical band with the period of $\approx310-320$ days \citep{2021AN....342..531Z}.  
The variable X-ray counterpart was detected by {\sl Swift}-XRT and {\sl XMM-Newton} \citep{Hinton2009,2011ApJ...737L..11B, 2012MNRAS.421.1103C,2018PASJ...70...61M}.  
A particularly extensive monitoring campaign carried out with {\sl Swift}-XRT. 
Analysis of these observations allowed for more precise measurements of the orbital period $(P_b=316.8^{+2.6}_{-1.4}$,  $P_b=313^{+11}_{-8}$ , and $317.3\pm0.7$  days  according to \citealt{2019AN....340..465M} [hereafter M+19],  \citealt{2018PASJ...70...61M}, and \citealt{2021arXiv210911894A}, respectively) and provided strong evidence for spectral variability as a function of the orbital phase.
According to \cite{2018PASJ...70...61M} fits to the X-ray spectra with an absorbed power-law (PL) model indicate that the object is more absorbed when it is brighter ($N_{\rm H}=(4.7\pm0.9)\times10^{21}$ cm$^{-2}$ vs.\ $(2.2\pm0.5)\times10^{21}$ cm$^{-2}$) and it  also, somewhat less significantly, experiences spectral hardening with the photon index changing from $\Gamma=1.54\pm0.12$ to $\Gamma=1.37\pm0.07$). 
M+19 analyzed the same {\sl Swift}-XRT data and also found a significant increase in  $N_{\rm H}$ near the peak of the light curve ($\phi=0.3-0.4$) as well as some evidence of photon index variability, with the spectrum 
 hardening  right before and right after the peak of the light curve. 
Recent {\sl NuSTAR} observations, confined to a narrow range of orbital phases ($\phi=0.1-0.2$),
confirm the PL nature of the spectrum and suggest that the slope is changing with the orbital phase \cite{2020ApJ...888..115A}.  
However, neither of these X-ray studies 
investigated the possibility of extended emission around  J0632. 
Such emission has been firmly detected around the HMGB LS~2883 \citep{2014ApJ...784..124K,2015ApJ...806..192P,2019ApJ...882...74H} and marginally detected around LS~5039 \citep{2011ApJ...735...58D}. 
On sub-arcsecond scales, extended radio emission has been reported from  J0632, LS~2883, LS~5039, LS~2883, and LS~I~+61 303 (Moldon et al.\ 2011).

 \begin{figure*}
\centering
\includegraphics[width=0.9\hsize,angle=0]{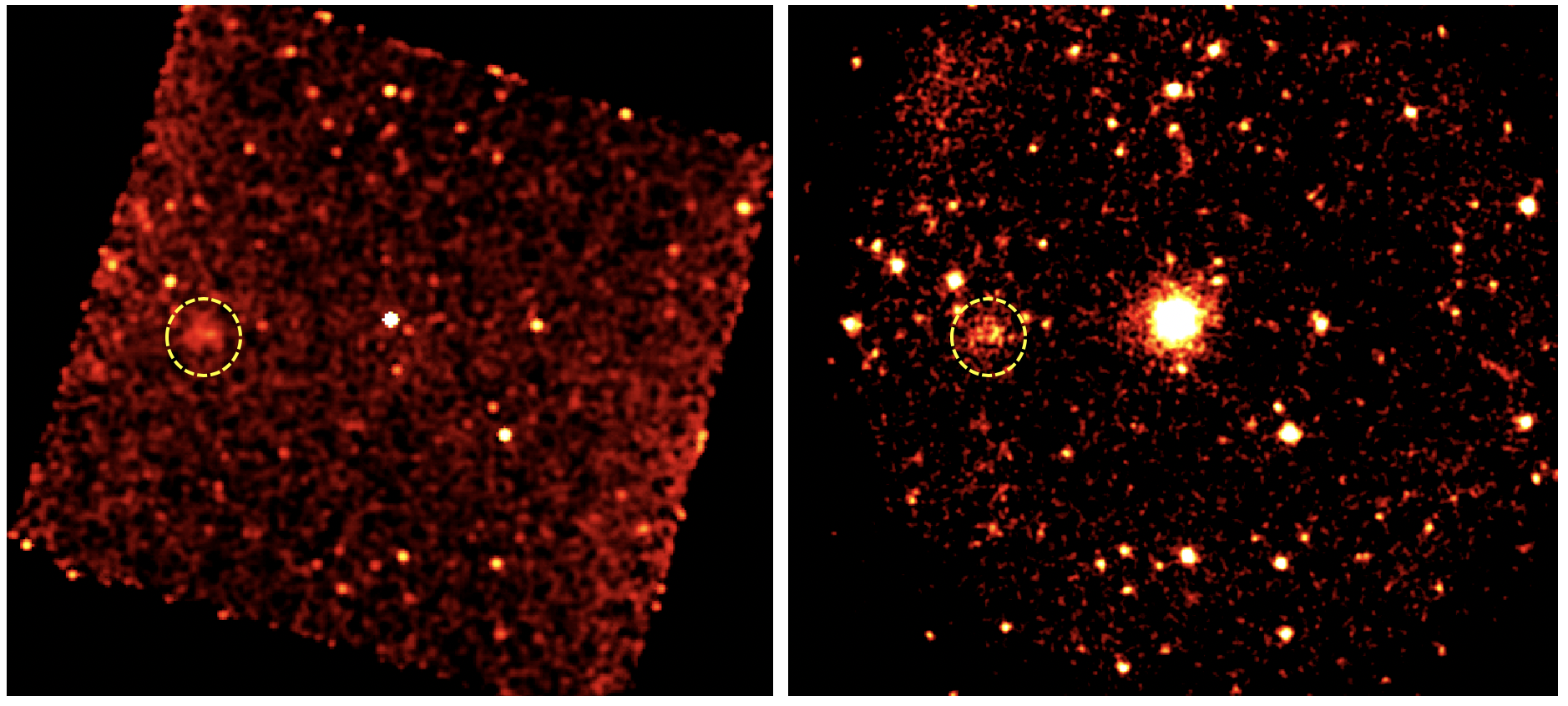}
\caption{\footnotesize
Combined  {\sl CXO}-ACIS-I (left; 0.7--8 keV,  152 ks) and {\sl Swift}-XRT (right; 0.3--10 keV, 967 ks) images. Both images are of the same angular size and show the same part of the sky. 
 }
\label{cxo-large}
\end{figure*}

The nature of the compact object in J0632 remains elusive despite the attempts to find pulsations in X-rays \citep{2011ApJ...737L..12R, 2020ApJ...888..115A}. 
An alternative way to constrain the compact object type could be the detection of extended emission in X-ray.
{\sl Chandra X-ray Observatory} ({\sl CXO}) observations of the most famous HMGB LS2883 (which hosts radio pulsar B1269--63) revealed a remarkably dynamic extended emission attributed to the interaction between the pulsar wind and the wind of the massive star (\citealt{2019ApJ...882...74H} and references therein).  
Detecting a similar extended emission in the case of J0632 would support the pulsar nature of the compact object in this system. 
This prompted us to carry out high angular resolution {\sl CXO}-ACIS observations of   J0632 and report the results here. 
In addition, we analyzed numerous archival {\sl Swift}-XRT observations which, once added up, result in a very long cumulative exposure of the J0632 field (967 ks) allowing for a very deep image sensitive to extended emission on larger scales, and providing a high-quality phase-resolved spectrum of the binary.  
On the other hand, the lack of extended emission, signatures of accretion activity (such as 
 flares, QPOs, or emission lines) would support a slowly rotating NS 
or a BH scenario for the compact object.   
To look for these signatures, we analyzed the archival data from  {\sl CXO}-ACIS
and {\sl Swift}-XRT.  
In Section 2 we describe the observations and the data reduction, while in Section 3 we describe the data analysis and report our findings. 
We discuss the implications of our results in Section 4, and conclude with a summary in Section 5.

\section{Data}
In our analysis described below we use four imaging observations performed with {\sl CXO}-ACIS between 2018-02-03 and 2018-02-21, and publicly available data from 275 {\sl Swift }-XRT monitoring observations taken over the period from 2009-01-26 -- 2018-12-14.
  
\subsection{CXO observations}
  
{\sl CXO} observed J0632 on four occasions during the two epochs (with two observations per epoch) separated by $\approx15$ days (see Table 1) in February 2018.  
During all observations the target was imaged on the ACIS-I array close to the aim point (offset was $\approx 50''$). 
The detector was operated in Very Faint mode with a time resolution of 3.2 s. 
The data were processed following standard procedures using CIAO v.4.11 and CALDB v.4.8.4.1. We also used {\sl Chart}\footnote{http://cxc.harvard.edu/ciao/PSFs/chart2/runchart.html} v.2 and MARX\footnote{https://space.mit.edu/cxc/marx/} v.5.4.0 to produce the observation-specific images of the {\sl CXO} PSF.
Due to the incorrectly implemented offset during the first two observations of J0632,
the source fell into the chip gap on ACIS-I array. 
This led to the loss of about 50\% of photons for J0632 and introduced artificial periodic modulation due the dithering of the target across the chip gap. 
After the problem was recognized, the observations were repeated without any issues (Epoch 2). We only use the second epoch data for the imaging analysis on arc-second scales (Section 3.1.2) as well as for the timing analysis (Section 3.2.1). 
However, for the large scale image (Section 3.1.1) and spectral (Section 3.3.1) analysis we use data from all 4 {\sl CXO} observations. 
The maximum expected pile-up fraction is $5\%$,
hence we neglect the impact of pile-up in our analysis below. 
  
We also included in our spectral analysis the only other archival 40 ks  {\sl CXO} observation that was obtained on 2011 February 13, with ACIS-S3 operated in Continuous Clocking (CC) mode (see \citealt{2011ApJ...737L..12R} for details). 
This observation provided more photons ($10,580$ net counts in 0.5-8 keV) than all imaging CXO observations taken together because the observation occurred near the 
 light curve peak
 when the target was bright
 and the background contribution to be negligibly small (2\% of total) even in  the CC mode.

\subsection{Swift-XRT observations}

J0632 has been frequently 
monitored by XRT between 2009 and 2018, with 275 observations taken during this period.  
We processed these data using the 
 {\sl Swift}-XRT data products generator
\footnote{https://www.swift.ac.uk/user\_objects/} hosted at the University of Leicester (the pipeline is described in detail by \citealt{Evans2009}). 
The data products include phase-resolved XRT spectra
and the merged image with exposure maps (with and without vignetting).

\subsection{Spectral Fitting}
  
We performed all spectral fitting using XSPEC v.12.11.1.  
For all fits to the binary's {\sl CXO}-ACIS spectra, we group them by 1 count per bin 
prior to fitting and use the {\tt cstat} implementation of Cash statistic available in XSPEC. 
For visualization purposes, all fitted spectra shown in the paper are re-binned in XSPEC using the ``setplot rebin'' command which does not impact the fitting.  
For the extended ``blob'' spectrum from {\sl CXO}, where the background contribution is substantial, we use larger grouping and and fit using the chi-squared statistic.  
For {\sl Swift}-XRT spectra, we used binning of 40 or more counts per spectral bin and chi-squared statistic for the fits.  
Unless specified otherwise, all uncertainties for the parameters inferred from spectral fits are given at 68\% confidence level 
for a single interesting parameter.

\section{Results}  
  
Below we describe in detail the results of imaging (Section 3.1), Timing (Section 3.2), and Spectral (3.3) analysis of the {\sl CXO}-ACIS and {\sl Swift}-XRT observations.

\subsection{CXO and Swift images}

Thanks to the long combined exposure taken by {\sl Swift}, both the XRT and ACIS images can provide useful information about arcminute-scale structures. 
The superb angular resolution of {\sl CXO} also allows for an arcsecond-scale investigation in the vicinity of J0632.

\subsubsection{Arcminute scales}

Figure \ref{cxo-large} (left panel) shows the combined ACIS-I image of the J0632 field (in 0.7--8 keV) after correction for the exposure map.  
The binary is the brightest source at the center of the ACIS-I field of view. 
The only region of {\sl extended} emission that stands out in the ACIS-I image is the ``blob'' (shown by the dashed yellow circle), with radius of $\sim35''$ 
located $\approx5.5'$ east of the binary
and centered at R.A.\  $\approx$6h33m21s and decl.\ $\approx+05^{\circ}47'42''$.

\begin{figure*}
\centering
\includegraphics[width=0.9\hsize,angle=0]{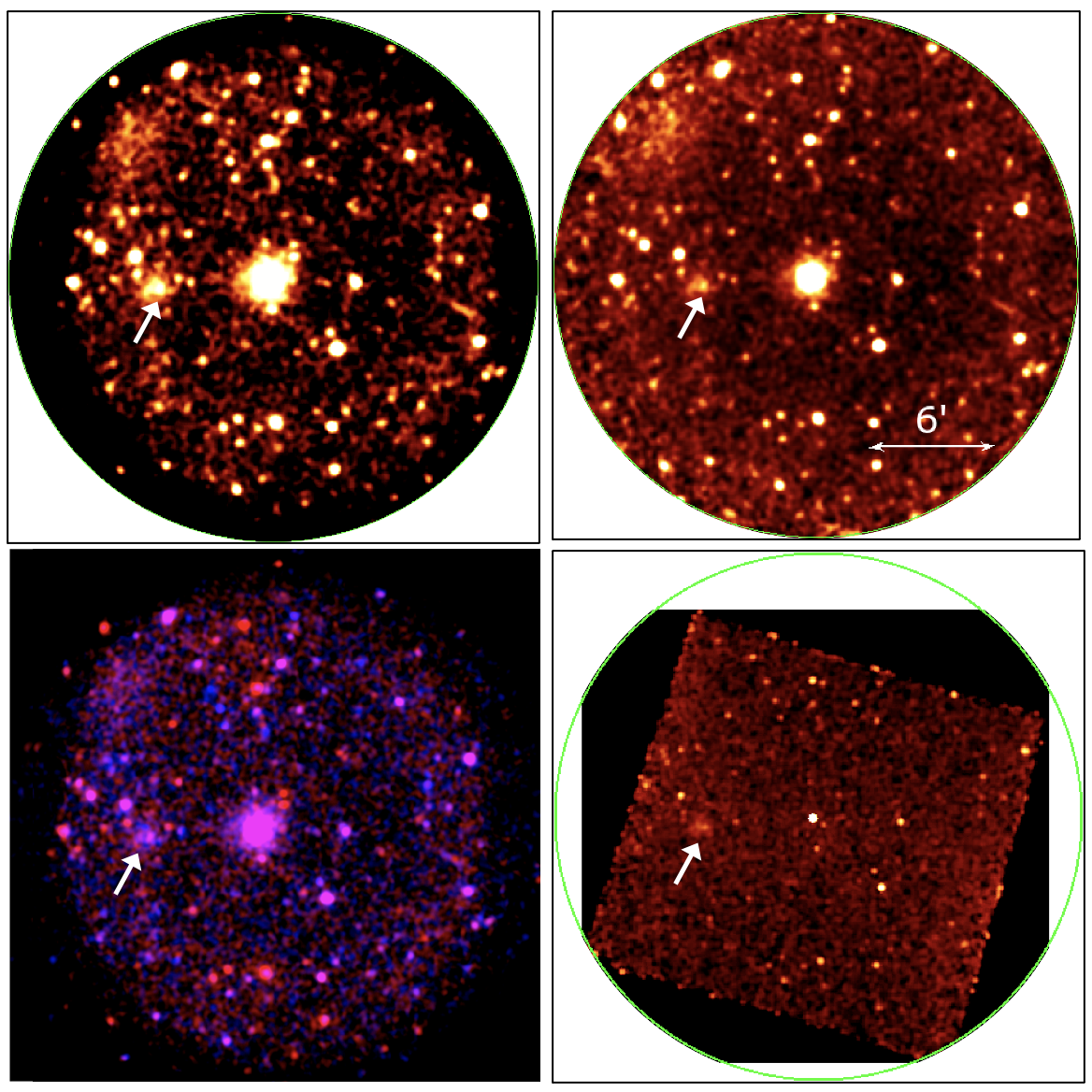}
\caption{\footnotesize
 The top left panel shows  967-ks    XRT image of the J0632 field. The same image corrected by the exposure map (with vignetting included) is shown on the top right.  The bottom right panel shows the same image as above but in false colors (0.3--2 keV -- red; 2--10 keV -- blue).  The exposure-corrected  {\sl CXO} ACIS image from Figure \ref{cxo-large} is shown in the bottom right panel. 
  All 3 images cover the same area on the sky. In all images North is up and East is to the left.
 }
\label{xrt-large}
\vspace{-0.0cm}
\end{figure*}

The merged 967-ks image produced from 275 XRT observations of the J0632 field and 
smoothed with an $r=7''$ Gaussian kernel is shown in the top left panel of Figure \ref{xrt-large}. 
The top right panel shows the same image divided by the exposure map (which accounts for both the varying exposure time and for the vignetting)\footnote{We used the UK Leicester University {\sl Swift}-XRT data products generator
to produce the merged image and the exposure map.}. 
The image in the bottom left panel is the same as above it but in false color showing hard emission in blue (2--10 keV) and soft in red (0.3--2 keV).  
The smaller {\sl CXO} ACIS-I image is shown for comparison in the right bottom panel.
In addition to the ``blob'' visible in the {\sl CXO} ACIS image, the XRT image shows a larger region of extended emission NE 
of the binary. 
The diffuse emission appears to be relatively hard, which can be due to its intrinsic spectrum or due to substantial intervening absorption which efficiently removes soft photons (see Section 3.3.1).

Interestingly, the exposure-corrected XRT image also suggest that HESS~J06232 is  either located inside a  ``cavity''  (region with a lower surface brightness) or surrounded by a  shell of brighter emission. 
We note that the cavity/shell is seen even {\sl before} the exposure map correction (top right panel) when the sensitivity is the highest at the center of the image due to the vignetting. 
Naturally, correcting for the vignetting by dividing by the exposure map further enhances the appearance of the cavity/shell.  
Unfortunately, the size of the XRT field of view is too small to determine reliably whether it is indeed a dimmer cavity within the larger region of enhanced diffuse emission (e.g., hot gas in young stellar cluster), or whether it is simply a region bounded by a 
 shell (e.g., due to an associated SNR). The analysis of the energy resolved images shows that the cavity/shell morphology is  pronounced  $<2$ keV 
 and barely noticeable $>2$ keV. 
The {\sl CXO}-ACIS-I 
 field-of-view is even smaller in size, and hence it is not helpful in investigating this large-scale structure. 
It is also not as deep as the XRT image, due to the much shorter exposure and the substantial sensitivity degradation\footnote{https://cxc.cfa.harvard.edu/ciao/why/acisqecontamN0013.html} $<1.5$ keV. 

\begin{figure*}
\centering
\includegraphics[width=0.77\hsize,angle=0]{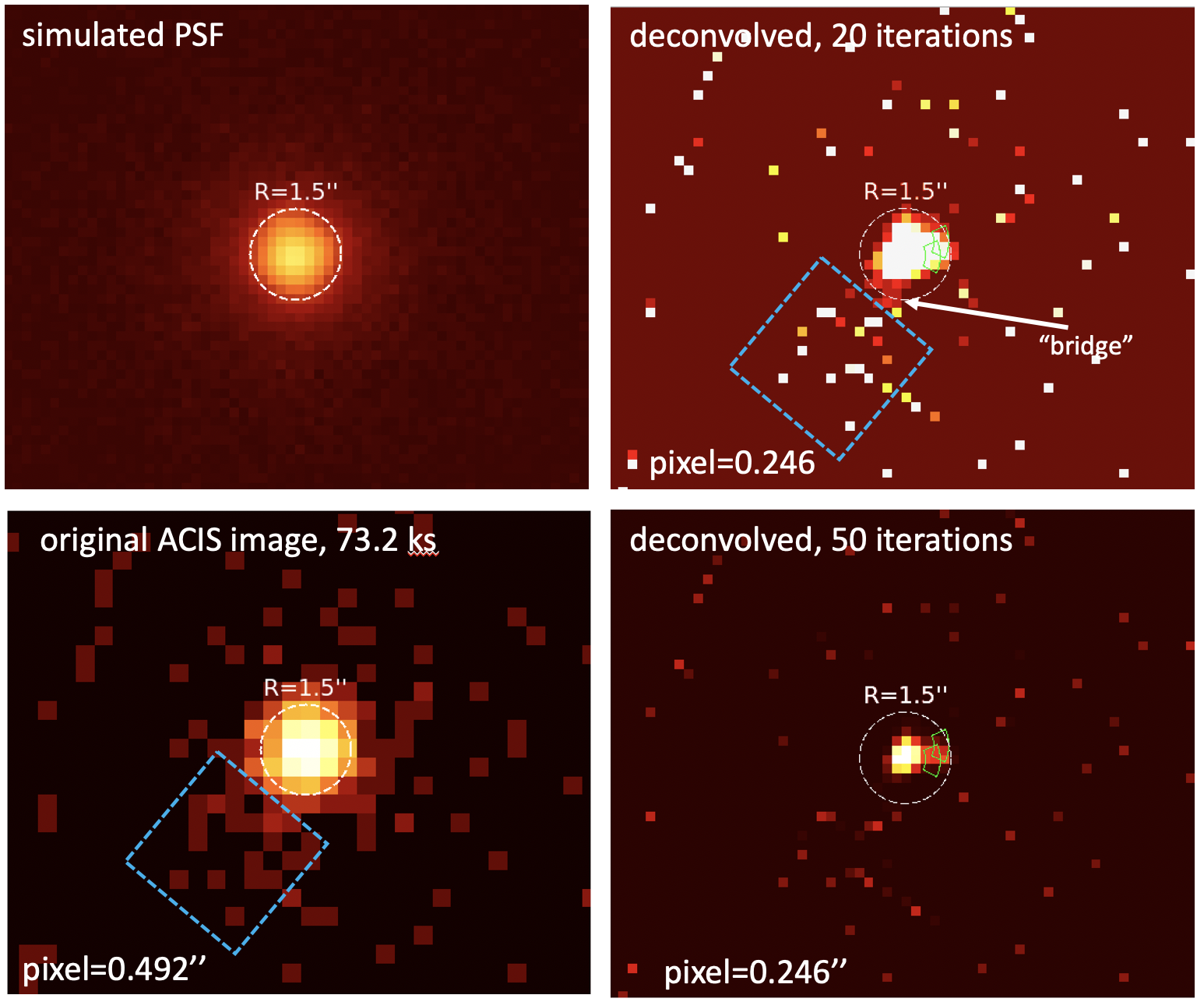}
\caption{\footnotesize
The vicinity of J0632 as seen in the 2nd epoch (see Table 1) {\sl CXO}-ACIS image (bottom left; 0.7-8 keV; 73.2 ks).  The top left panel shows PSF image simulated with the {\tt Chart} ray-tracing application and projected onto the detector plane using MARX.
The top and bottom right images show deconvolved (with the simulated PSF) ACIS images at sub-pixel resolution obtained with {\tt arestore} using 20 and 50 iterations, respectively.  
}
\label{cxo-small}
\end{figure*}

\subsubsection{Arcsecond scales}

As the next step we performed careful evaluation of emission on smaller angular scales in the vicinity of the binary. 
Figure \ref{cxo-small} shows a zoomed-in view of the binary obtained by combining two ACIS images from Epoch 2. 
Since these two observations are less than a day apart and the target is placed at nearly the same position with respect to the optical axis of the telescope
 (with the roll angles being  nearly the same), we do not expect any issues in the combined image of J0632's vicinity caused by  non-uniformities on the chip or varying PSF.  
We also made sure that the images are well-aligned (using the field sources to check the alignment) before co-adding them.  
Even at the native ACIS resolution (bottom left panel in Fig.~\ref{cxo-small}) there appears to be some excess of photons SE of the binary (within the blue dashed square). 
This excess seems to be present out to about $3''$ from the binary but it is only significant at $\simeq2\sigma$ level\footnote{There are 24 photons within the dashed rectangle shown in Figure \ref{cxo-small}. 
If the same rectangle is placed on the other 3 sides the numbers of photons are 9, 14, and 15 (the average is 12.7 and standard deviation is 3).}.

To look for the extended emission on even smaller scales, we followed the CIAO/Chart/MARX recipe\footnote{http://cxc.harvard.edu/ciao/threads/psf.html} to simulate the observation-specific PSF (top left panel in Fig.~\ref{cxo-small}) and used it with {\tt arestore}\footnote{http://cxc.harvard.edu/ciao/ahelp/arestore.html} to produce the deconvolved images with sub-pixel resolution (right panels in Fig.~\ref{cxo-small}). 
The resulting image (deconvolved with 20 {\tt arestore} iterations) in the top right panel shows a hint of ``bridge'' of emission in the direction of the above-mentioned excess (within the blue dashed square).  
The image also revealed the artifact caused by the known {\sl Chandra} mirror defect\footnote{http://cxc.harvard.edu/ciao/caveats/psf\_artifact.html} whose position angle is quite different from that of the above-mentioned asymmetric excess.  
At 50 iterations (bottom right panel) the point source looks more concentrated  but the asymmetric excess emission disappears. 
The latter  
is not surprising because the 
 deconvolution procedure does not conserve the number of photons.

\begin{figure*}
\centering
\includegraphics[width=0.99\hsize,angle=0]{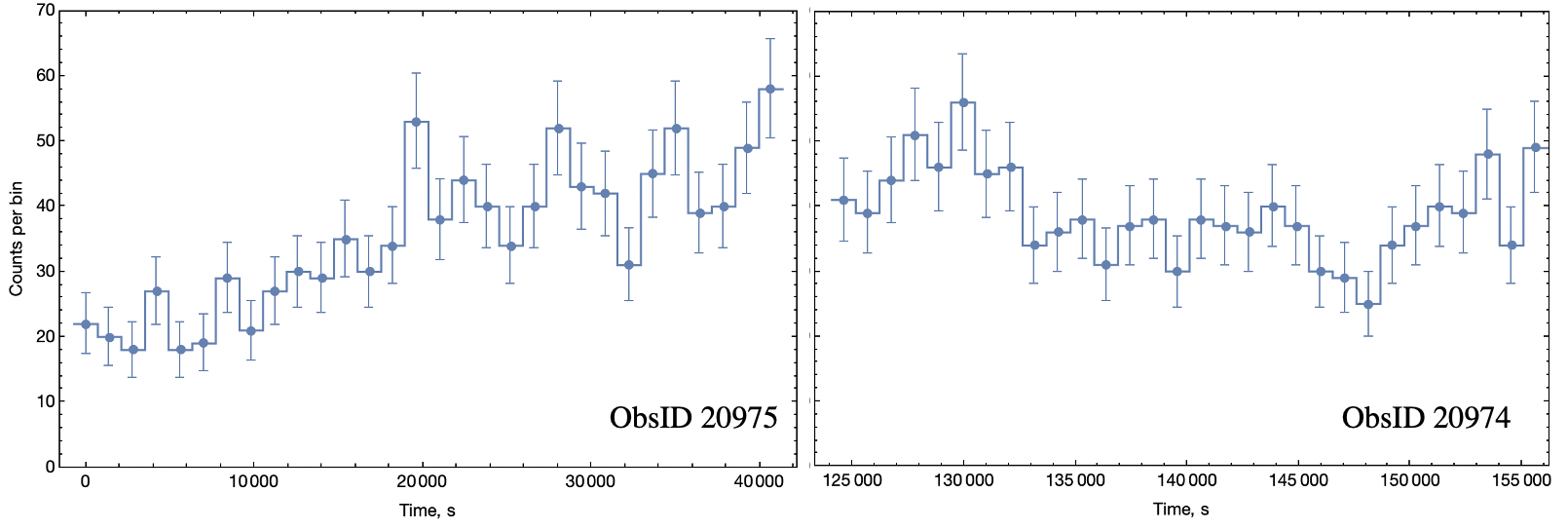}
\includegraphics[width=0.99\hsize,angle=0]{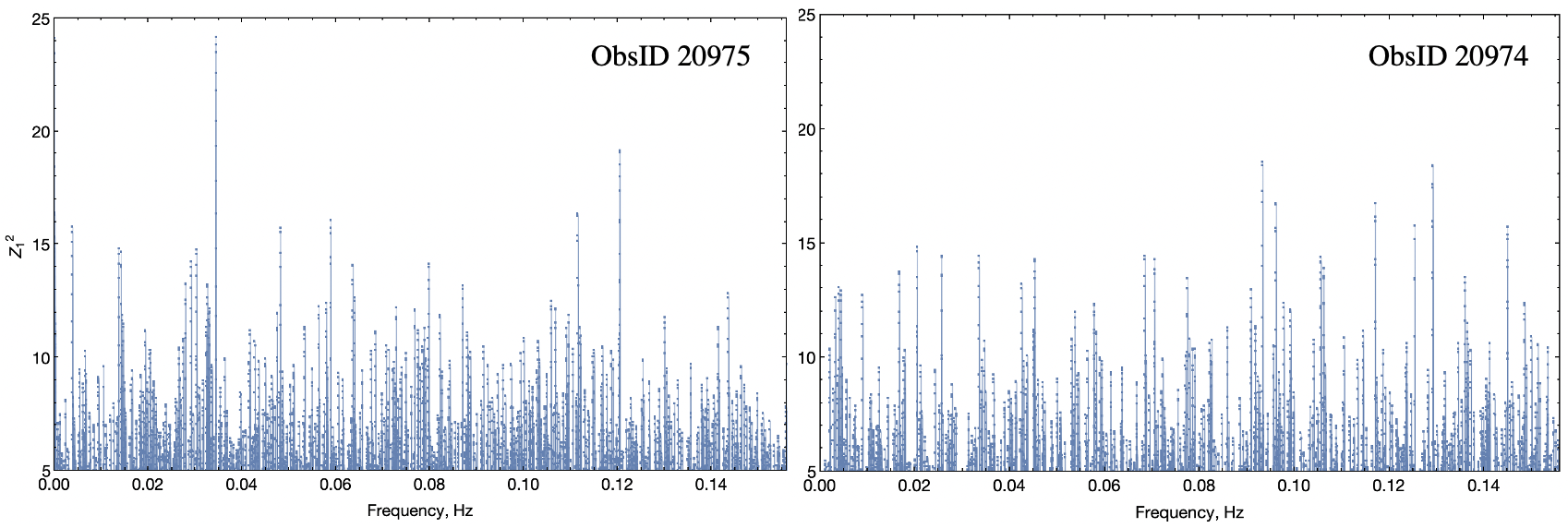}
\caption{\footnotesize
Lightcurves (top) and  $Z_1^{2}$  (bottom)   for two {\sl CXO} observations from Epoch 2.  
}
\label{cxo-z1}
\end{figure*}

\subsection{Timing}

Several different orbital  ephemerides  have been suggested for J0632 (e.g., \citealt{2012MNRAS.421.1103C, 2018PASJ...70...61M, 2019AN....340..465M}). 
There is currently no consensus on which ephemeris is the most accurate, and all suggested  ephemerides have substantial uncertainties. 

\subsubsection{CXO}

Across these $30-40$ ks observations, the J0632 count rate varues by a factor of 2 which suggests that spectral properties may also vary on these or shorter timescales (see Section 3.3.1). 
However, the number of photons collected by {\sl CXO} does not allow us to meaningfully probe  spectral changes on shorter timescales. 
More sensitive {\sl XMM-Newton} observations must be undertaken to quantify the spectral variability on shorter timescales.
Figure \ref{cxo-z1} shows the light curves\footnote{We do not show light curves from the Epoch 1 observations because they are affected by the 
dithering across the chip gap.} of two {\sl CXO} observations from Epoch 2 with $t=0$ corresponding to the start time of the first observation (in both cases).

To search for possible periodic signals in the Epoch 2 data we first corrected the arrival times of all photons to the solar system barycenter using the CIAO tool {\tt axbary}. 
Because of the uncertain binary ephemeris, we do not attempt to correct the photon arrival times for the binary motion (R\"omer delay; \citealt{1976ApJ...205..580B}) while looking for a periodic signal in the ACIS-I data. 
The lack of an accurate orbital ephemerides represents a significant obstacle affecting the sensitivity of the previous searches for pulsations at high enough frequencies  \citep{2011ApJ...737L..12R,2020ApJ...888..115A}. 
However, the limited 3.2-s resolution of imaging {\sl CXO} observations only permits a search for periodic pulsations up to $f_{\rm max}=0.1562$ Hz. 
This frequency is sufficiently low for us not to worry about the R\"omer delay correction. 
Indeed, the acceleration velocity of the compact object in J0632 does not exceed $a_{\rm max}\approx10$ cm s$^{-2}$  (see, e.g., equation B9 in \citealt{2021ApJ...915...61V}) which is attained for the ephemeris from \cite{2012MNRAS.421.1103C}. 
We 
searched for pulsations separately in each observation from Epoch 2 as well as  jointly in both observations. 
For the joint search with total time span $T_{\rm span}=1.807$ days (or 156,140 s), the native frequency resolution is $ \delta f=1/T_{\rm span}=6.4$ $\mu$Hz. 
The maximum shift in frequency around frequency $f$ during the time interval $T_{\rm span}$ due to the binary motion can be estimated as  $a_{\rm max} T_{\rm span} f/c\approx50 (f/1~{\rm Hz})$~$\mu$Hz . 
Therefore, the maximum possible shift is  $\approx8$~$\mu$Hz  for $f<f_{\rm max}=0.1562$ Hz. This shift is 
comparable to the native frequency resolution $\delta f=6.4$ $\mu$Hz (the characteristic width of the peak in the Fourier power spectrum). Therefore, the signal should not be washed out substantially even for frequencies close to $f_{\rm max}=0.1562$ Hz. 
It is even more so for the individual observations because their time span is even smaller.

For {\sl CXO} observations from Epoch 2  we calculated the 
Rayleigh statistic ($Z_n^{2}$ for n=1,2; \citealt{1983A&A...128..245B}) on the grid of $N=10 \times f_{\rm max} T_{\rm span}$  frequencies uniformly distributed between $f_{\rm min}=50$~$\mu$Hz  and $f_{\rm max}=0.1562$ Hz. 
The grid provides a factor of 10 better resolution compared to $T_{\rm span} f_{\rm max}$ thus ensuring that the $Z_n^{2}$ peak values are accurately captured.  
The highest value, $Z_{\rm max,1}^2=24.2$ at $f_{\rm max, 0}\approx0.03442$ Hz,
was found for ObsID 20975. 
However, the significance of this value is only 96.35\% after accounting for $T_{\rm span} f_{\rm max}=6559$ trials frequencies. 
This maximum  value of $Z_{1}^2$ is substantially smaller for the ObsID 20974 data. 
We also calculated $Z_2^{2}$ for both observations taken together but the significance of the maximum $Z_2^{2}$ value turned out to be smaller  than 96\%. 
Therefore, we 
do not find  statistically significant periodic signal in the data.

To estimate the upper limits on the intrinsic pulsed fraction (the background correction is negligible)  for  periods in the range of  6.4 s to 20 ks from  ObsID 20975  we solved equation (8) of \cite{1995ApJ...447..807F}  and found the limits to be 27\%, 30\% and 31\%  at the confidence levels of 95\%, 99\%, 99.7\%, respectively.  These limits are comparable to 32\%--48\% (in 0.005--8 Hz; at 99.7\% confidence) obtained  from {\sl XMM-Newton} observation obtained near $\phi=0.43$ and analyzed by \cite{2011ApJ...737L..12R}.

\subsubsection{Swift}

Since each XRT observation collected relatively few photons and the observations were spread over many years we did not attempt to search for a periodic signal from the compact object.  
We only use XRT arrival times to produce the binary light curve and assign binary phases to the data.
Zero phase, $\phi=0$, was set to be %
$T_0=54857$ MJD 
and the binary period used for the folding was  $P_{\rm} =316.8$ days as reported in M+19 who  analyzed a similar amount of {\sl Swift}-XRT data from the archive. 
All phases given below rely on the M+19 ephemeris, unless specified otherwise. 

\begin{figure*}
\centering
\includegraphics[width=0.5\hsize,angle=0]{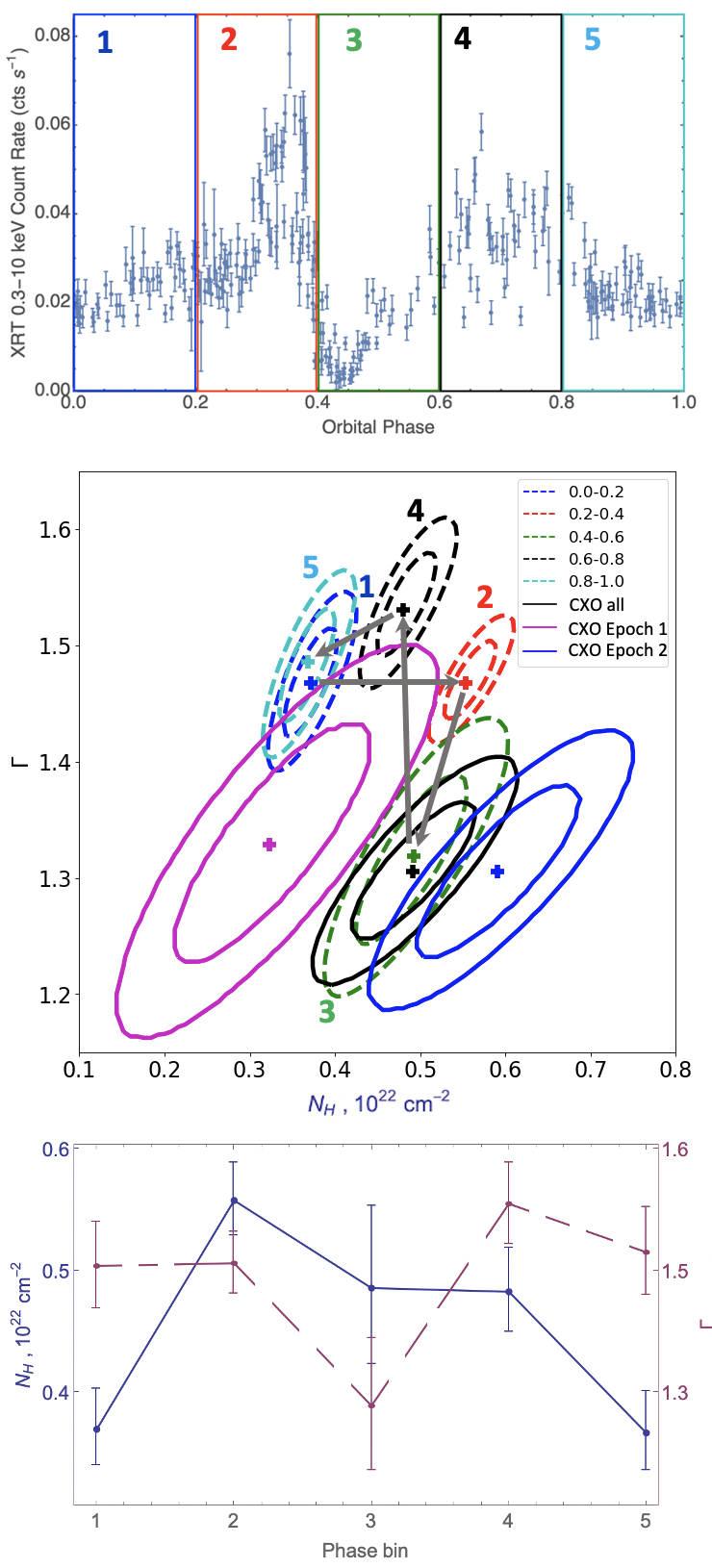}
\caption{\footnotesize
The top panel shows the average  count rate (for each of the 275  XRT observations) as a function of the binary phase assuming $P_{\rm} =316.8$ days (M+19) and zero phase ($\phi=0$) set to be at $T_0=54857$ MJD. 
The five phase intervals shown by colored boxes are used to extract spectra discussed in Section 3.3.2. 
The middle panel shows, in the $N_{\rm H}-\Gamma$ plane, confidence contours (1$\sigma$ and 2$\sigma$ levels are shown)
for the XRT spectra (dashed contours)
extracted from the data corresponding to the 5 bins shown in the XRT light curve (top panel), and 3 ACIS confidence contours (solid contours) extracted from Epoch 1 ($\phi=4.0$) and Epoch 2 ($\phi=4.5-4.6$) data fitted individually (magenta and blue solid lines) and fitted jointly (black solid line).
The arrows connecting the best-fit values from the 5 XRT spectra track the evolution with increasing phase bin number. 
See Sections 3.3.1 and 3.3.2 for details. 
The bottom panel is a simpler (but less informative) way to show the evolution of $N_{\rm H}$ (solid black line) and $\Gamma$ (dashed purple line) with the binary phase as seen from the XRT observations. 
 }
\label{xrt_plots}
\vspace{-0.0cm}
\end{figure*}

Figure \ref{xrt_plots} (top panel) shows the J0632 light curve where each datapoint corresponds to average count rate in a single XRT observation. 
We use the light curve to define 5 equally-sized intervals (0.2 in phase) motivated by the light curve morphology, from which we extract and fit spectra in Section 3.3.2. 
The XRT data used in our analysis span approximately 10 binary cycles. 
There are some moderate changes in the shape of the light curve (particularly near the two light curve maxima) between the binary cycles. 
These changes can be seen from Figure 2 in M+19, but we 
use our own multi-cycle light curve shape for defining the 5 broad phase bins.  
We also note that the light curve shown in Figure 2 of M+19 plots flux versus the binary phase. 
The conversion from from count rate to flux is subject to uncertainty due to impossibility of determining the spectrum from a single XRT observation which, on average, contains only $\sim75$ photons from the source. 
Therefore, if the spectrum changes (which it does as we show below)
and the conversion is based on the averaged spectrum, the resulting fluxes will be inaccurate.

\subsection{Spectra}

\subsubsection{CXO}

Below we fit the {\sl CXO} spectra of J0632 from the imaging and continuous clocking (CC) mode observations as well as the spectrum of the  resolved extended ``blob'' located $5.5'$ east of the binary using imaging observations only.

\begin{figure*}
\centering
\includegraphics[width=0.9\hsize,angle=0]{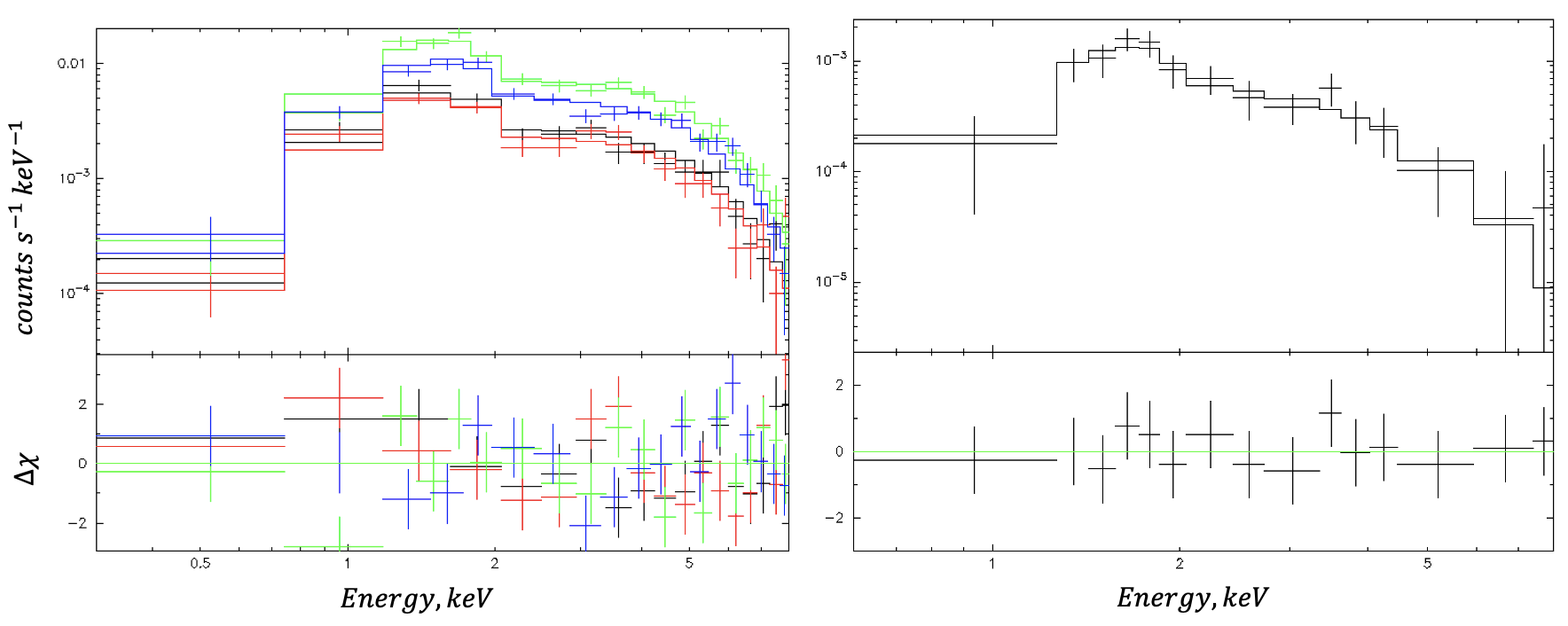}
\caption{\footnotesize
The ACIS spectra of J0632 (left) and the extended blob (right) extracted from all 4 {\sl CXO} observations and fitted with an absorbed PL model. Although in both cases the spectra were fitted simultaneously, for the blob we co added all 4 spectra for visualization purposes because it is fainter and not expected to vary ith time.  For the fits to the binary spectra the normalizations are untied while $\Gamma$ and $H_{\rm H}$ are tied together (see Section 3.3.1 for details).
}
\label{cxo-4spec}
\end{figure*}

{\em Imaging  observations:}
For J0632 we extracted the spectra from all 4 observations using an $r=3''$ circle centered on the binary using CIAO's {\tt specextract} script.  
Since the effective area calculation takes into account the reduction of sensitivity due to the dithering across the ACIS-I chip gap, we included in our analysis both Epoch 1 and 2 data.
There are a total of 3,168 photons within the extraction aperture from all 4 data sets taken together. 
Since the source is bright, the background contribution within the extraction aperture is very small and can be neglected for all energies. 
We first fitted all four spectra of J0632 (see Figure \ref{cxo-4spec}, left) simultaneously with an absorbed PL model where normalizations were untied (they are clearly varying between the 4 observations) while the photon index, $\Gamma$, and the absorption Hydrogen column density, $N_{\rm H}$, were tied together. 
The fit quality was good with the reduced $\chi_{\nu}^2$=0.994 for $\nu=$86 degrees of freedom (d.o.f). 
The best-fit PL has $\Gamma=1.32\pm0.06$ and $H_{\rm H}=(5.2\pm0.7)\times10^{21}$ cm$^{-2}$.  
The average absorbed flux in 0.5-8 keV is $4.9^{+0.2}_{-0.1}\times10^{-13}$ erg cm$^{-2}$ s$^{-1}$.

The two epochs of {\sl CXO} observations differ in binary phase only slightly, by $\approx\Delta\phi=0.05$. 
Both observations took place right after $\phi=0.4$ where there is a sharp transition from light curve maximum to the light curve minimum (see below).  
One 
can expect to see spectral variability associated with the transition. 
Therefore, we separately fitted spectra from Epoch 1 and Epoch 2 with an absorbed PL model. 
For each epoch (consisting of two observations) the spectra were fitted simultaneously with normalizations being untied (due to the variability on small timescales; see Figure \ref{cxo-z1}) and the rest of the fitting parameters being tied together. 
For Epoch 1 we obtained $N_{\rm H}=(3.4\pm1.1)\times10^{21}$ cm$^{-2}$, $\Gamma=1.3\pm0.1$, and fluxes of $(4.8_{-0.4}^{+0.2})\times10^{-13}$ erg cm$^{-2}$ s$^{-1}$ and $(4.1_{-0.2}^{+0.3})\times10^{-13}$ erg cm$^{-2}$ s$^{-1}$  for the first and second observations\footnote{The values are different since the model normalizations were untied.}, respectively.  
For Epoch 2 we obtained $N_{\rm H}=(5.9\pm0.9)\times10^{21}$ 
cm$^{-2}$, $\Gamma=1.30\pm0.07$, and fluxes of $(4.5_{-0.1}^{+0.2})\times10^{-13}$ erg cm$^{-2}$ s$^{-1}$ and $(6.6\pm0.2)\times10^{-13}$ erg cm$^{-2}$ s$^{-1}$ for the first and second observations. 
These results indicate changes in the  hydrogen column density and flux, but no measurable changes in the spectral slope. 
Since the uncertainties of $N_{\rm H}$ and $\Gamma$ are strongly correlated, the differences are best represented by the confidence contours shown in Figure \ref{xrt_plots}. 
Note that the values of fluxes indicate that the observations took place at the lowest (faintest) state, in agreement with the expectations for these
 phases  (cf.\ ``fluxed'' light curve from {\sl Swift}-XRT shown in the left panel of Figure 2 in M+19).

We also extracted the spectrum of the ``blob'' (Figure \ref{cxo-4spec}, right panel) from an $r=35''$ region centered on the blob. 
In this case we extracted the background spectrum from a source-free circular ($r=70''$) region located on the same chip. 
The total number of counts from all 4 observations is 1,229, of which only $\approx$30\% are from the ``blob'' while the rest are from the background. 
We fitted the spectra from all four observations simultaneously with an absorbed PL model.  
For visualization (only) purposes, we co-added spectra from all 4 observations (and their responses) and show the total spectrum together with the best-fit model (based on the  simultaneous fit to all 4 spectra).  
The fit quality is good with  $\chi_{\nu}^2=$0.77 for $\nu=$21 d.o.f. 
The best-fit $N_{\rm H}=(1.9^{+0.8}_{-0.6})\times10^{22}$ cm$^{-2}$ and $\Gamma=2.6^{+0.6}_{-0.5}$. 
We also fitted the spectrum with a thermal plasma (mekal) model\footnote{A thermal Bremsstrahlung model gives very similar parameters and quality of the fit.} and obtained an equally good fit with $N_{\rm H}=(1.4\pm0.5)\times10^{22}$ cm$^{-2}$ and $kT=3_{-1}^{+2}$ keV.

{\em CC-mode  observation:} We also fitted the spectrum of J0632 from the ACIS CC-mode observation that occurred near $\phi\approx 0.36$.
The source spectrum 
was extracted from a $3''$-long box while the background was taken from two  $9''$-long boxes flanking the source region (with $1''$ gaps between the source and background regions). 
Prior to fitting, we grouped the source spectrum by a minimum of 50 counts per bin and fitted it with an absorbed PL model (see Figure \ref{xrt_cxo_cc}, top panel).  
The fit quality was good with the reduced $\chi_{\nu}^2$=0.99 for $\nu=$156 d.o.f. 
The best-fit PL model has $\Gamma=1.63\pm0.03$ and $n_{\rm H}=(5.8\pm0.3)\times10^{21}$ cm$^{-2}$.  
The corresponding absorbed flux in 0.5-8 keV is $2.78^{-0.04}_{+0.02}\times10^{-12}$ erg s$^{-1}$.

\subsubsection{Swift}

We also perform both phase-integrated and phase-resolved fits to the {\sl Swift}-XRT spectra and compared the results to those from the {\sl CXO} spectral fits. 
We note that, unlike M+19, we used a factor of 2 larger phase bins defined by the global changes in the light curve shape to increase the statistics and sensitivity to spectral changes, if the latter are correlated with the light curve morphology.

\

\begin{figure*}
\centering
\includegraphics[width=0.54\hsize,angle=0]{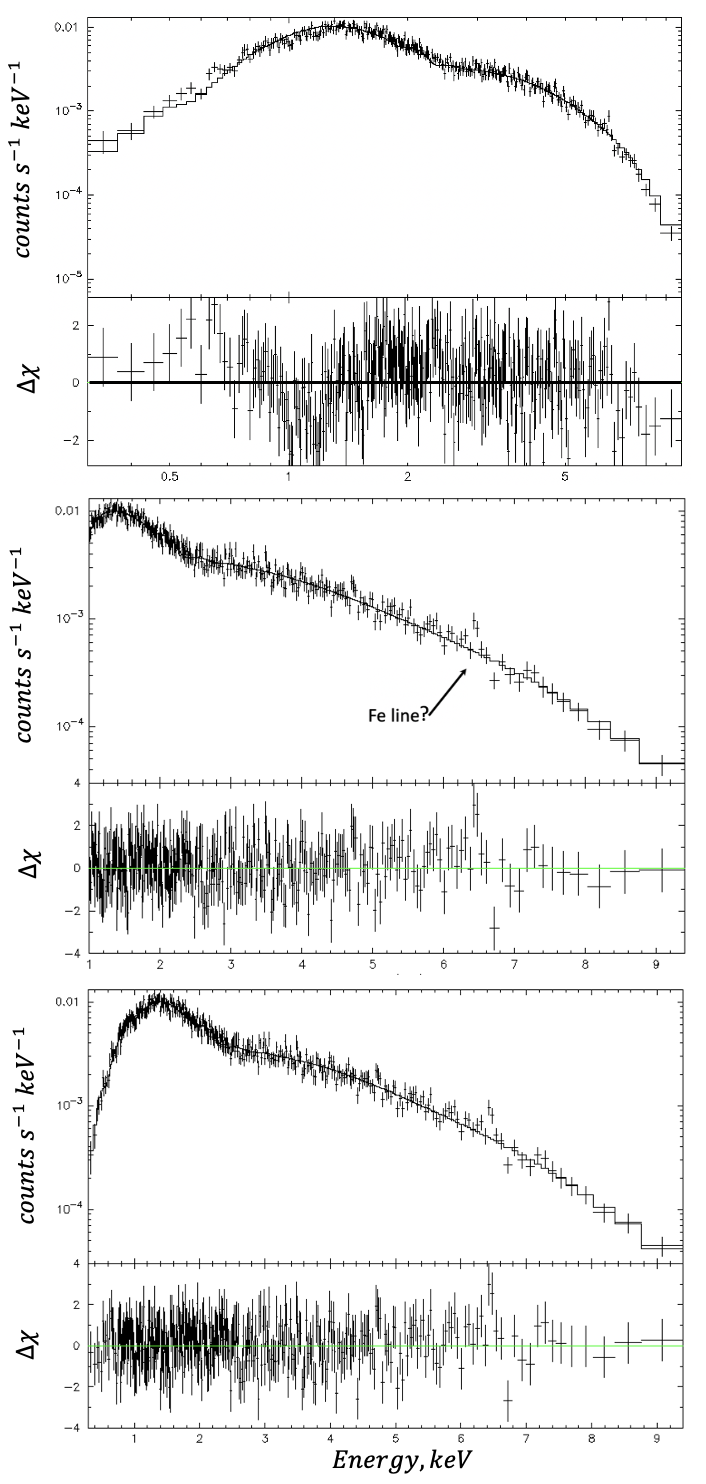}
\caption{\footnotesize
Combined XRT spectrum of J0632  fitted with  an absorbed PL ({\tt tbabs*pow}) model in two  energy ranges 0.3-10 keV (top panel) and 1-10 keV (middle panel), and with an absorbed thermal Bremsstrahlung model with a Gaussian absorption line (bottom panel) in the 0.3-10 keV range. 
}
\label{xrt_spec_merged}
\vspace{-0.3cm}
\end{figure*}

{\em Merged spectrum from all XRT observations:} The XRT spectra were extracted 
from the 275 XRT observations and combined into a single spectrum using the Swift-XRT data products generator\footnote{The pipeline picks an extraction radius based on the source's count rate in each observation.  In most of the XRT observations the count rate 
in the $0.01-0.05$ cts s$^{-1}$ range, so an $r=15$ pixel ($r=35.4''$) extraction aperture was used.  See \citet{Evans2009} for details.}.  
The combined spectrum contains 20,889 photons (cf.\ 3,168 in all 4 ACIS-I observations). 
We binned the spectrum requiring a minimum of 50 counts per bin and fitted it in the 0.3-10 keV energy range.  
The background (taken from a source-free annulus around the source) contribution is 
only $\approx3\%$ in 0.3-10 keV, and, hence its impact on the fits is negligible except in the highest and lowest energy bins. 
As can be seen from Figure \ref{xrt_spec_merged}, the fit with an absorbed PL model (best-fit $n_{\rm H}=0.47\pm0.02$ cm$^{-2}$ and $\Gamma=1.47\pm0.02$)
is bad ($\chi_\nu^{2}=1.25$ for $\nu=297$)
due to the large systematic residuals around 1 keV and the less substantial residuals between 6.4-6.8 keV.
However, if the fit is restricted to $>1$ keV (Fig.~\ref{xrt_spec_merged}; right panel) the spectrum can be adequately described ($\chi_{\nu}^2=$1.00 for $\nu=$263) by an absorbed PL model with $\Gamma=1.67\pm0.03$ and $N_{\rm H}=(7.8\pm0.4)\times10^{21}$ cm$^{-2}$ with the only noticeable residuals localized around 6.4 keV.  
Alternatively, the entire 0.3-10 keV spectrum fits well ($\chi_\nu^{2}=0.97$ for $\nu=294$) by an absorbed PL with $N_{\rm H}=(4.5\pm0.2)\times10^{21}$ cm$^{-2}$, $\Gamma=1.58\pm0.03$ and a Gaussian absorption line centered at $E_{\rm gabs}=1.09_{-0.03}^{+0.04}$ keV with width $\sigma_{\rm gabs}=243_{-39}^{+46}$ eV. 
Note that for the fits in the 0.3-10 keV range, the value of $N_{\rm H}$ does not depend on whether the {\tt gabs} component is included or not because the $N_{\rm H}$ value is largely determined by the spectrum at energies below the absorption feature. 
The value of $\Gamma$ is only slightly affected by the inclusion of the {\tt gabs} component.
The same spectrum can be described equally well ($\chi_\nu^{2}=0.98$) by a  Bremsstrahlung\footnote{or high temperature thermal plasma, e.g., {\tt apec}.} model with $N_{\rm H}=(3.9\pm0.1)\times10^{21}$ cm$^{-2}$, $kT=14\pm1$ keV, normalization $N=(2.95\pm0.03)\times10^{-4}$ cm$^{-5}$, and absorption line with $E_{\rm gabs}=1.09\pm0.03$ keV and $\sigma_{\rm gabs}=165_{-41}^{+49}$ eV.  
Note that all there 
fits still exhibit well-localized residuals between 6.4 and 6.8 keV. The larger $N_{\rm H}$ value for the PL fit restricted to $>1$ keV energies can be explained by the increase in  $N_{\rm H}$ accounting ro the residuals around 1 keV which does not happen when the lower energies are included.

\begin{figure*}
\centering
\includegraphics[width=0.45\hsize,angle=0]{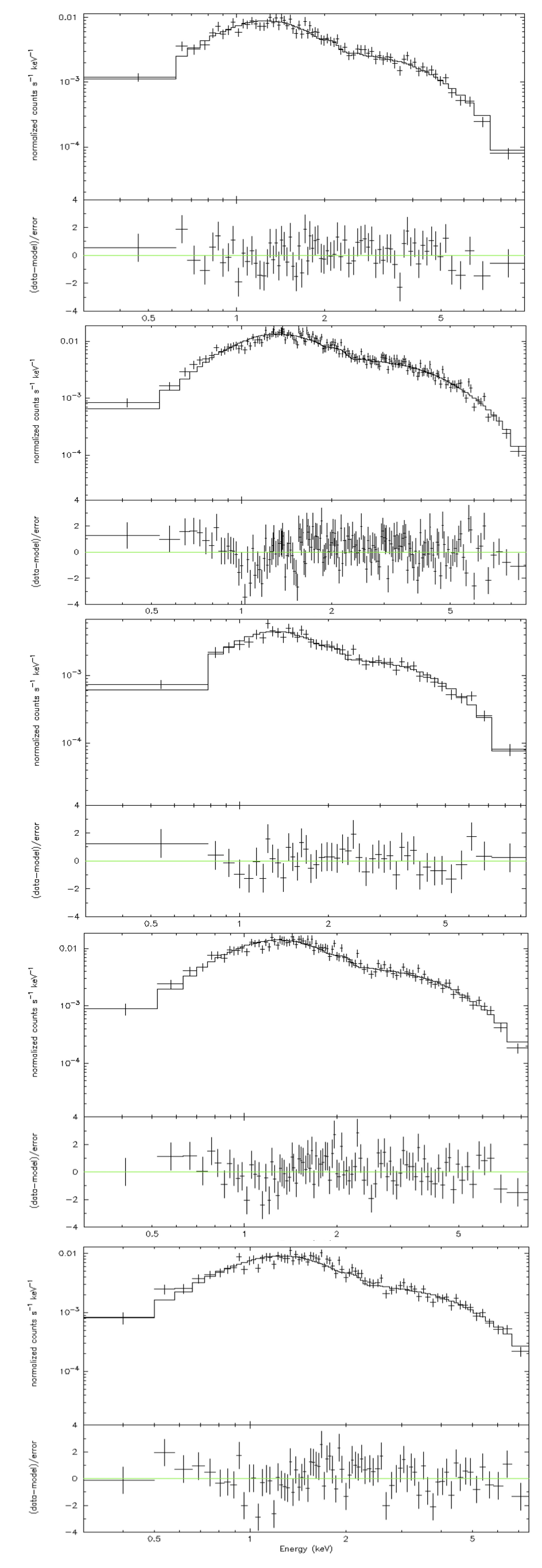}
\caption{\footnotesize
XRT spectra for phase bins 1-5 (from top to bottom) fitted with an absorbed PL model.
}
\label{xrt_spec_phase}
\vspace{-0.3cm}
\end{figure*}

\begin{figure*}
\centering
\includegraphics[width=0.45\hsize,angle=0]{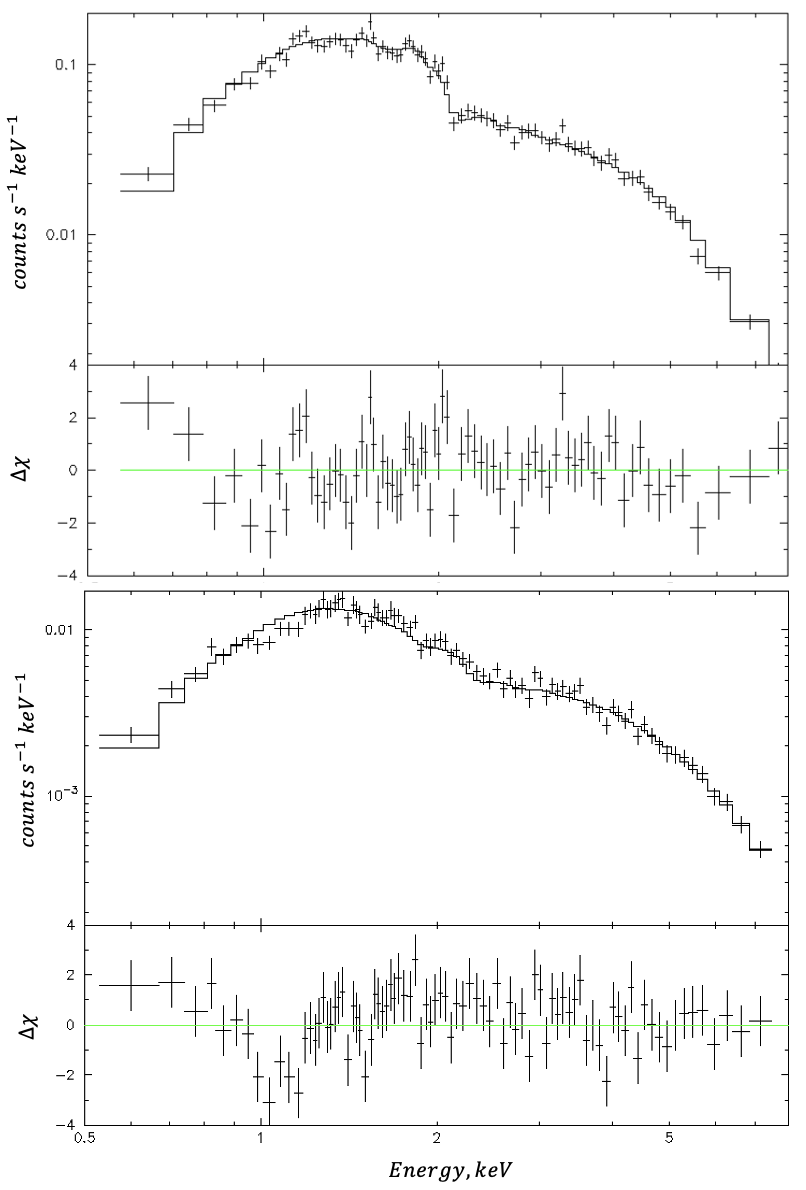}
\caption{\footnotesize
Comparison of the ACIS CC mode observation spectrum ($\phi=0.36$; top)  with the XRT spectrum for the $\phi=0.2-0.4$ bin (bottom).
}
\label{xrt_cxo_cc}
\vspace{-0.3cm}
\end{figure*}

{\em Phase-resolved spectra:} Intrigued by the residuals seen in the absorbed PL model fits to the merged spectrum, we  explored the phase-resolved spectra to see if the magnitude of these residuals varies with the binary phase. 
The large residuals around $1$ keV (for the simple absorbed PL model) could be the result of combining spectra that vary with time (e.g., the circumbinary contribution to $N_{\rm H}$ may vary).
We combined the spectra 
from 
 observations belonging to each of the five phase bins (shown in Figure \ref{xrt_plots}, top panel) 
into five merged phase-resolved spectra using the same Swift-XRT data products generator.
The phase-resolved spectra fitted with an absorbed PL model are shown in Figure  \ref{xrt_spec_phase}. 
The best-fit values of $N_{\rm H}$ and $\Gamma$, and the corresponding confidence contours  are plotted in Figure \ref{xrt_plots}. 
The contours imply varying $N_{\rm H}$ and somewhat less significant changes in $\Gamma$. 
The XRT values of $N_{\rm H}$  and $\Gamma$ for the phase bin 3 are consistent with those from the joint fit to all 4 {\sl CXO}-ACIS spectra.  
For simplicity, we do not include the ${\tt gabs}$ component while determining $N_{\rm H}$  and $\Gamma$. 
We do not expect these values to change substantially if ${\tt gabs}$ was included, based on 
what we saw for the phase-integrated spectra (see above).
The putative 6.4 keV feature from the phase-averaged XRT spectrum is not noticeable in the phase-resolved spectra, which suggests that the photons contributing to the weak feature in the combined spectrum are not concentrated in one of the 5 phase bins. 
Conversely, the 
 systematic residuals near 1 keV to be the strongest in the 2nd phase bin (which includes the light curve peak) with the weaker residuals 
  seen in the spectra from some other phase bins (4 and 5). 
Therefore, we conclude that the large residuals around $1$ keV are not likely to be of instrumental origin. 
Another observation that was taken near the light curve maximum is the {\sl CXO}-ACIS CC-mode observation that occurred at at $\phi=0.36$. 
Interestingly, an absorbed PL model fit to this spectrum (see Figure \ref{xrt_cxo_cc}) also shows some residuals near 1 keV but they are not  exactly the same as in the XRT spectrum from the 2nd phase bin ($\phi=0.2-0.4$).

\section{Discussion}

J0632 is one of the few known Galactic HMGBs for most of which the nature of the compact object remains elusive. 
Pulsars were only identified in LS~2883/B1259--63 and TeV 2032+4130 which have substantially longer orbital periods (3.5 and $\approx50$ years, respectively) compared to J0632. 
In both cases, extended emission has been resolved with {\sl CXO} \citep{2019ApJ...882...74H,2020arXiv200616022A}. 
In the case of TeV 2032+4130, this emission is seen on larger angular scales and is likely to be nonvariable (at least over a decade time scale). 
The arcminute-scale emission can be attributed to the PWN powered by the young PSR J2032+4127, which is only 
very infrequently perturbed by the passage of a massive star due to the very elliptical orbit. On the other hand, for a much more compact LS 5039 HMGB with a 4-day orbital period, arcimunte-size X-ray emission has been reported by \cite{2011ApJ...735...58D}  while \cite{2007ApJ...664L..39P} found tentative ($3.2\sigma$)  evidence for a more compact, $\simeq12''-24''$ in diameter, faint extended emission in the 50-ks {\sl CXO} observation of HMGB LS I +61 303 having  orbital period of 26.5 days.

For B1259, most of the arcsecond-scale extended emission is strongly variable.  
The brightest part of it are the episodic ejections in the direction of the binary apastron  likely launched near the time of periastron passage. 
The ejected clouds, which most likely represent shocked and compressed pulsar wind (with a possible admixture of stellar wind), appear to accelerate and expand as they move from the binary. 
The ejected blobs are not always visible because they fade and/or dissipate sooner than the next one is ejected during the subsequent periastron passage. Moreover, it appears that the ejecta clumps vary in intensity and sometimes may not be discernible while still contributing to a fainter asymmetric diffuse emission reported first by \citet{2011ApJ...730....2P}  (Figure 1, bottom panel, in that paper). 
This faint asymmetric emission is also oriented in the direction of apastron and may represent  the remains of multiple smaller and slower moving ejections. 

For the only other Galactic HMGB with a known pulsar, TeV J2032+4130 (PSR J2032+4127), the faint arcminute-scale diffuse emission was first noticed in the 50-ks {\sl CXO}-ACIS observation.  
Recently, \citet{2020arXiv200616022A} re-analyzed 166-ks of {\sl CXO} data from multiple observations and showed the combined image (in their Figure 1) where the hard diffuse emission  surrounding TeV J2032+4130 is seen much more clearly. 
This hard spectrum supports the PWN interpretation of this emission which must be powered by PSR J2032+4127.

\begin{figure*}
\centering
\includegraphics[width=0.9\hsize,angle=0]{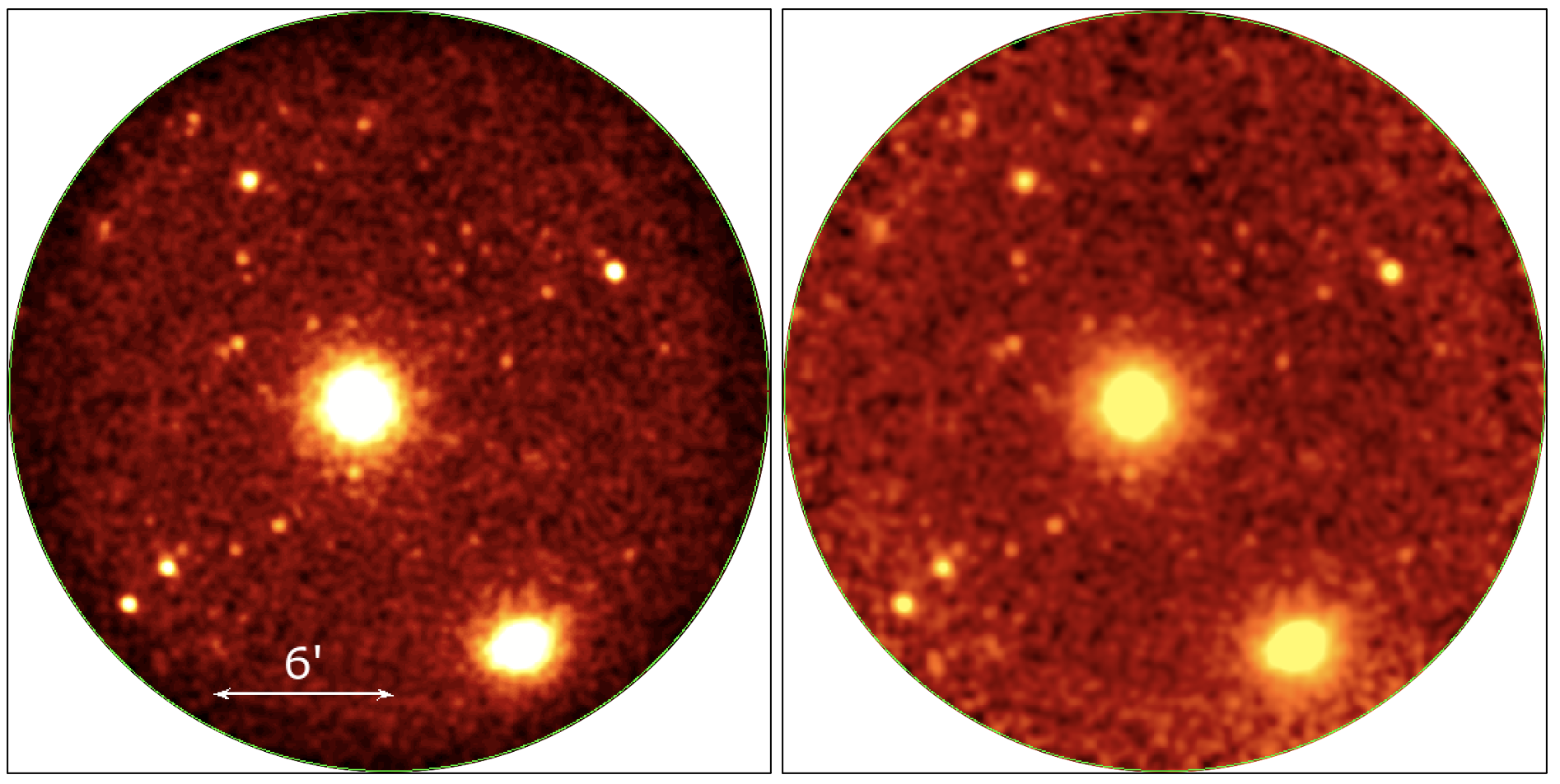}
\caption{\footnotesize
The left panel shows a 310-ks XRT image of the B1259--63 field  (the HMGB is the bright source near the center). The same image corrected by the exposure map (with vignetting included) is shown on the  right.  Both images cover the same area on the sky.
}
\label{xrt-b1259}
\vspace{-0.0cm}
\end{figure*}

The {\sl CXO} images of J0632 hint at an asymmetric emission, albeit at a lower significance than for B1259, which may also be extending in the direction of periastron (see Figure 7 in \citealt{2018PASJ...70...61M}). 
Unfortunately, the emission is too faint to say anything definitive about its spectrum or morphology and the direction to periastron is not certain (see Figure 2 in \citealt{2020ApJ...888..115A}).  
We note that for B1259--63, the faint asymmetric emission becomes apparent only in deep ACIS exposures. 
Therefore, a deeper ACIS imaging of J0632 may be warranted.

The evidence for the shell/cavity around J0632 is intriguing. If confirmed with observations having larger field-of-view, it could represent the host SNR of J0632. For the radius of $\approx10'$  (see Figure \ref{xrt-large}) the nominal Sedov age estimate gives $t\approx830 (d/1.5~\rm{kpc})^{5/2}(n/E_{51})^{1/2}$ yrs where 
$n$ cm$^{-3}$ is the local ISM density 
  and $10^{51}E_{51}$ erg 
  is  the SN explosion energy.
 
 As one can see, young rotational pulsar of this age is excluded, because it would be much more luminous than observed. It means either the shell-like structure is not associated an SNR forward shock,  the distance  or the density is much larger than assumed, or the compact  object is not a typical rotation-powered pulsar \citep{2008AIPC..983..171K}.

We note that in the case of B1259, where the age of the compact object is $\sim330$ kyrs (the pulsar spin down-age), no extended emission was previously reported on arcminute scales. To look for such emission we have co-added 164 archival {\sl Swift} XRT observations\footnote{Although there are many {\sl CXO} ACIS observations of B1259--63, only a small subarray was activated  during those observations making then not suitable for searching for extended emission on large angular scales.} of B1259--63 (310 ks of exposure in total) in a way similar to how we did it for J0632. 
The resulting images, shown in Figure \ref{xrt-b1259}, do not reveal any obvious extended emission on arcminute scales nor they show a shell/cavity similar to what we see for J0632 (cf.\ Figure \ref{xrt-large}). The lack of the SNR would not be surprising in the case of B1259 if the spin-down age is a fair estimate of the true age because the host SNR of PSR B1259-63 would have dissipated at the age of $\sim$300 kyrs.
The non-detection of the shell/cavity in B1259 also helps to conclude that the shell/cavity structure seen in the deep XRT image of J0632 is unlikely to be caused by instrumental effects.

\begin{figure*}
\centering
\includegraphics[width=0.97\hsize,angle=0]{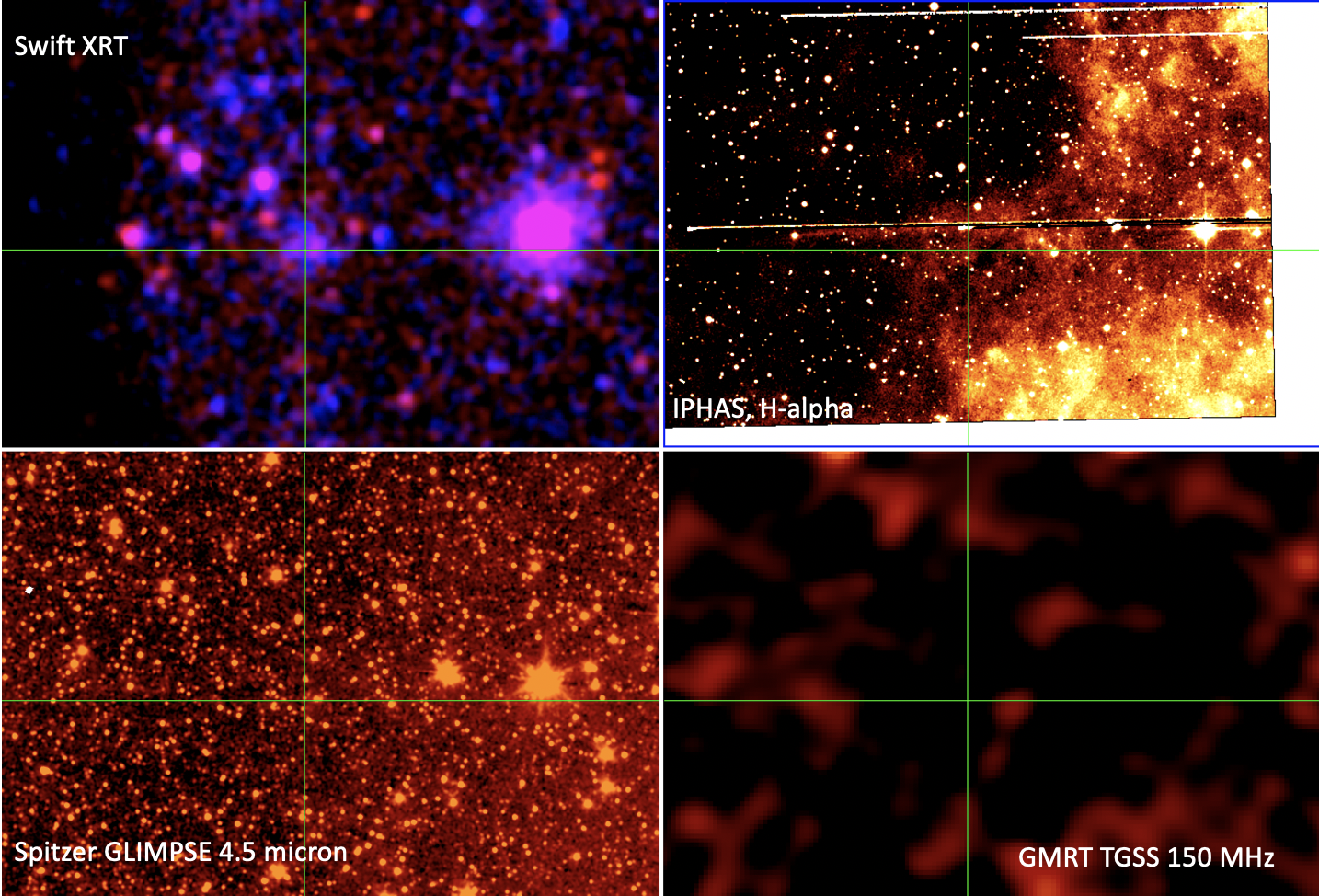}
\caption{\footnotesize
Multiwavelength images centered on the blob. The extended X-ray emission does not appear to have a counterpart in IR, H$_{\alpha}$, or radio.  All four images show the same part of the sky. 
}
\label{blob_mw}
\end{figure*}

In addition to the shell/cavity structure with non-uniform enhancements, the only other arcminute-scale structure resolved both in ACIS and XRT images is the puzzling blob located  $\approx5.5'$ east of the pulsar. 
The blob did not change its position over the period of time of 9 years covered by {\sl Swift} and {\sl CXO} observations, which precludes motion with speed $\gtrsim0.05c$ that would have resulted in a noticeable, $\approx20''$, shift during 9 years (cf.\ the clumps ejected from  B1259 which are moving with an average velocity of $\approx0.07c$; \citealt{2019ApJ...882...74H}).  
The spectrum of the blob, 
is 
relatively hard and can be 
described by an absorbed PL with  $\Gamma\simeq2.6$ and $N_{\rm H}\simeq1.9\times 10^{22}$ cm$^{-2}$, or by a thermal plasma emission with $kT\simeq3$ keV and $N_{\rm H}\simeq1.4\times 10^{22}$ cm$^{-2}$. 
These values are rather uncertain, but $N_{\rm H}$ appears to substantially larger than that for J0632, thus disfavoring the relation between the two. The spectra of the clumps ejected from B1259 are harder, with $\Gamma=1.2-1.9$ \citep{2019ApJ...882...74H}.
However, similar to the clumps in B1259, the blob does not appear to have a counterpart at lower wavelengths (see Figure \ref{blob_mw}) including the {\sl Spitzer}-IRAC images where a young cluster or a star-forming region would be apparent. 
If the shell/cavity structure seen in the  deep XRT image of J0632 is indeed associated with the host SNR, then  the blob could be also part of the SNR structure but the $N_{\rm H}$ for the blob would still be unexplained. 

The recent report of possible magnetar-like periodicity in LS~5039 \citep{2020PhRvL.125k1103Y}  prompted us to search for pulsations in ACIS data up to 0.156 Hz, the maximum frequency  afforded by the ACIS time resolution in the imaging mode.
The strongest evidence of periodicity found, at  $\approx29.05$ s, is  only significant at the 96\% confidence level, which is too low to confidently claim a detection. 
The limiting maximum frequency of our search  (0.156 Hz) is low enough to neglect the R\"omer delay.   \cite{2011ApJ...737L..12R} performed searches up to 180 Hz near the light curve maximum using the {\sl CXO}-ACIS CC-mode observation and up to 8 Hz using {\sl XMM-Newton}-EPIC with negative results. 
However, without a R\"omer delay correction, which requires knowing an accurate binary ephemeris, 
it is difficult to find the periodic signal at high frequencies using a simple periodogram approach.
Therefore, the non-detection of periodicity does not rule out a young pulsar nature of the compact object in J0632. 
Once a more accurate binary ephemeris is obtained, it would be worthwhile to revisit the  {\sl CXO}-ACIS CC-mode observation to see if pulsations can be detected. 
In the absence of an accurate ephemeris, a more promising approach to the detection of possible pulsations would be collecting more photons within shorter time interval(s) for which the binary motion correction is  unimportant. This could be done 
with {\sl XMM-Newton} (in Small Window mode) or with {\sl NICER} (given the lack of bright nearby objects). 
 
Thanks to the 
multiple {\sl CXO} observations, we are able to explore the spectral variability on short timescales. 
Despite having both ACIS epochs during the flux minimum state (since they both show similar average fluxes corresponding to that of the minimum state according to XRT measurements; see left panel of Figure 2 in M+19), we observed a difference in $N_{\rm H}$ values between the two epochs separated by 15-18 days. 
This may be caused by the observer's line-of-sight passing through an inhomogeneous circumbinary medium (e.g., clumpy wind of the massive star).
If so, one could, in principle, see variation on even smaller timescales if larger number of photons is collected using with {\sl XMM-Newton}  or  {\sl NICER}. 
For the clumpy wind scenario, measuring variability timescale(s) for $N_{\rm H}$  could put a limit on the size(s) of the Be-star wind clumps that are farther away from the star than the compact object (for an accreting BH or NS scenario) or the intrabinary shock (for the young pulsar scenario). In the latter case the characteristic size of inhomogeneities can be estimated as  $8.6\times10^{11}(v_{\rm psr}/100~{\rm km/s})(\tau/1~{\rm day})$ cm, which is the distance traveled by the pulsar near the periastron in one day scaled to the  velocity expected for the putative pulsar in J0632 near periastron. Therefore, in the $N_H$ variability on timescale of a week caused by the clumpy wind near periastron, then the size of the clump would be roughly about 10 times larger than the star radius at the distance (from the massive star) where the clump meets the pulsar.  Since the clumps are expected to expand as they travel away from the massive star this size may be reasonable. Moreover, variations in $N_H$ may be occurring on smaller timescales which cannot be probed with the existing data quality and then the estimate given here is rather an upper limit on the clump size. 

To characterize the variability on larger timescales and check if the {\sl CXO} observations near the light curve minimum result in fluxes which are in agreement with previous measurements, we extracted XRT spectra from 5 spectral bins.  
One of these bins ($\phi=0.4-0.6$) encloses the phases of the {\sl CXO} observations. 
We found that, for the combined spectrum of all XRT observations within that phase bin, both the best-fit $\Gamma$ and $N_{\rm H}$ are in good agreement with those inferred from the 
{\sl CXO} spectrum (see Figure \ref{xrt_plots}). 
The agreement suggests that although $N_{\rm H}$ does exhibit noticeable fluctuations on a smaller timescales, these fluctuations can average out because the value indicated by the XRT spectral fit is based on multiple XRT observations from multiple binary periods. 
The evolution of J0632's spectral parameters with the phase suggest that $N_{\rm H}$ increases significantly (almost by a factor of 2) during the light curve maximum, stays fairly large (on average) during the light curve minimum, and then returns to a lower value for the rest of the orbit. 
The spectrum marginally hardens during the light curve minimum but for other 4 phase bins the slope of the spectrum remains approximately the same. 
The large change in $N_{\rm H}$ could be naturally explained by increased obscuration near the  periastron, when the compact object might be passing through a decretion disk of the Be star (see Figure 4 in  M+19; also \citealt{2021arXiv210911894A}). 
We, however, note that the ephemerides calculated by both \cite{2012MNRAS.421.1103C} and \cite{2018PASJ...70...61M} place the periastron at very different phases from the light curve maximum (see \citealt{2020ApJ...888..115A}). 
If one of these ephemerides is correct, then a different explanation for the increase in $N_{\rm H}$ during the light curve maximum may be needed. 
Guided by their numerical simulations and the ephemeris from \cite{2012MNRAS.421.1103C},  \cite{2017MNRAS.471L.150B} suggest that the sudden drop after the lightcurve maximum is associated with the disruption (in the apastron direction) of the surrounding dense stellar wind shell by the accumulating energetic pulsar wind. This would support having periastron near the lightcurve minimum. Finally, we note that the existing data support correlation between the X-ray flux and $\Gamma$ unlike the anti-correlation seen in LS 5039 \citep{2021ApJ...915...61V}. This result is in agreement with the  findings reported by M+19 and \cite{2018PASJ...70...61M} for J0632.

The phase-resolved spectra (see Figure \ref{xrt_spec_phase}) show the substantial 
phase-dependent 
residuals (for either the absorbed PL fit) centered around 1 keV and a weak Fe K$\alpha$ line near 6.4 keV which does not seem to exhibit strong dependence on the binary phase. 
The 6.4 keV line is only visible in the merged spectrum, likely, due to its weakness. 
Therefore, if the line is 
real, it is likely to be present in the emission associated with the wide range of orbital phases. 
Although the presence of the 6.4 keV line is frequently associated with the accretion disk in XRBs, it may be also produced by photo-ionization of cold and dense Be wind material in the colliding wind region. 
However, in this case, the line should be stronger when the pulsar is within (or close to) the decretion disk (assuming its wind does not disrupt it completely). 
Alternatively, if the pulsar is not very powerful, some residual accretion may be happening even if the pulsar wind stops most of the infalling Be star wind material.  
  
The origin of the large residuals near 1-keV is rather 
puzzling regardless of whether the X-ray emission comes from the intra-binary shock, shocked pulsar wind, or accretion disk/column/jet near the compact object (if it is not a young pulsar). 
The 
 residuals can simply be the artifact of combining 
spectra with $N_{\rm H}$ (and other spectral parameters) varying over a fairly broad range of phases ($\phi =0.2-0.4$) since it is difficult to attribute them to the absorption (e.g., from Fe L complex) because Fe K$\alpha$ is seen in the emission (albeit it is weak).  
Although the ACIS CC-mode spectrum (Figure \ref{xrt_cxo_cc}, top panel), obtained in a single observation taken in the very narrow range of phases $\phi=0.362\pm0.001$ exhibits some residuals near 1 keV they, do not well match those seen in the XRT spectrum from the much broader (but overlapping) $\phi =0.2-0.4$ range.  
We also tried to fit the XRT spectrum with thermal plasma models (mekal and apec) and found that a high temperature, $kT=10-13$ keV, is needed to describe its overall shape. 
At such  temperatures, the optically-thin thermal plasma spectrum is very close to the Bremsstrahlung spectrum shown in Figure \ref{xrt_spec_merged}. 
The spectral feature around 1 keV is, of course, still there. 
Formally, the feature can be satisfactorily described by a broad Gaussian absorption line ({\tt gabs} model in XSPEC) which, however, does not shed light onto its origin. A more sensitive observation with {\sl XMM-Newton} or {\sl NICER}, which collects enough photons within a short interval of time, may shed light on the origin of these systematic residuals.

\section{Conclusions}

Although the {\sl CXO}-ACIS images revealed hints of extended emission in the vicinity of J0632 on arcsecond scales, it is too faint to be credibly interpreted as similar to that detected in the vicinity of the B1259-63 HMGB.  
We did not find any pulsations in the ACIS data up to 0.156 Hz. 
Therefore, the nature of the compact object in J0632 remains uncertain. 
The {\sl Swift}-XRT images suggest the existence of a cavity/shell around the binary with a radius of about $\sim7'$ (for cavity) or $\sim10'$ (for shell) corresponding to a physical size of 2.6 at 4.3 pc at the assumed J0632's distance of 1.5 kpc. 
If confirmed by observations with larger field-of-view (e.g., with {\sl XMM-Newton}), the shell could belong to the host SNR of J0632.
The significant variability in $N_{\rm H}$ appears to be happening on timescales as short as 15-18 days (according to the ACIS data) and may be attributed to the changing circumbinary environment along the observer's line-of-sight. 
The substantial variability in $N_{\rm H}$ (albeit on longer time scales) is also supported by our analysis of XRT data.  
This variability may be particularly large near binary phases encompassing the light curve peak, but more sensitive observations with a higher cadence are needed in that phase interval to confirm this finding. 
The analysis of XRT data also revealed a puzzling and apparently phase-dependent spectral feature near $1$ keV, which is strongest during the phase interval of the light curve maximum.
The combined XRT spectrum also shows hints of a weak 6.4 keV Fe K$\alpha$ line whose strength does not appear to noticeably vary with the binary phase.

\vspace{+0.5cm}

{\em Facilities:} \facility{{\sl CXO} (ACIS)}, \facility{VLA}, \facility{GMRT},
\facility{{\sl Swift} (XRT)},
\facility{{\sl Spitzer}}


\section*{Acknowledgements}

Support for this work was provided by the National Aeronautics and
Space Administration through {\sl Chandra} Award Number GO8-19040X
issued by the {\sl Chandra} X-ray Observatory Center, which is
operated by the Smithsonian Astrophysical Observatory for and on
behalf of the National Aeronautics and Space Administration under
contract NAS8-03060.  JH acknowledges support from an appointment to the NASA Postdoctoral
Program at the Goddard Space Flight Center, administered by the Universities Space Research Association
under contract with NASA.  We are grateful to George Pavlov for  helpful discussions. We also thank the anonymous referee for  the useful suggestions that helped us to improve the paper. 


\bibliographystyle{aa}

\end{document}